\documentclass[prb, preprint, a4paper, superscriptaddress]{revtex4}

\usepackage{amssymb}
\usepackage{bm}% bold math
\usepackage{epsfig}
\usepackage{graphicx}
\usepackage{amsmath}

\def\d2{\mbox{$d^2I/dV^2$}}

\begin{document}

\title{Inelastic effects in electron transport studied with
wave packet propagation.}
\author{S. Monturet$^1$ and N. Lorente}
\affiliation{Laboratoire Collisions, Agr\'egats, R\'eactivit\'e,
UMR5589, Universit\'e Paul Sabatier,
118 route de Narbonne, 31062 Toulouse c\'edex, France}
\affiliation{
 Centre d'Investigaci\'o en Nanoci\`encia i Nanotecnologia (CSIC-ICN), \\
Campus de la UAB, 08193 Bellaterra, Spain}

\date{\today}

\begin{abstract}
A time-dependent
approach is used to explore inelastic effects during electron transport
through  few-level systems. We study a  tight-binding
chain with one and two sites connected to vibrations. This simple
but transparent model gives
insight about inelastic effects, their meaning and the approximations
currently used to treat them. 
Our time-dependent approach allows us to trace back the
time sequence of vibrational excitation and electronic interference,
the vibrationally introduced time delay and the electronic phase shift.
We explore a full range of parameters
going from weak to strong electron-vibration coupling, from tunneling
to contact, from one-vibration description to the need of including 
all vibrations for a correct description of inelastic effects in
transport. We explore the validity of single-site resonant models as well
as
its extension to more sites via molecular orbitals and 
the conditions under which multi-orbital,
multi-vibrational descriptions cannot be simplified. We explain the
physical meaning of the spectral features in the second derivative
of the electron current with respect to the bias voltage. This permits
us to nuance the meaning of the energy value of dips and peaks.
Finally, we show that finite-band effects lead to electron back-scattering
off the molecular vibrations in the regime of high-conductance,
although the drop in conductance at the vibrational
threshold is rather due to the rapid
variation of the vibronic density of states.
\end{abstract}
\maketitle

\section{Introduction}

The importance of inelastic effects in electronic transport in molecular
junctions is widely recognized and it is a rich, active research field.
Several recent reviews on the topic give a clear idea of 
its breadth~\cite{Galperin2008,Galperin2007}. 
Advanced new experimental techniques show that
electrons transport through a few-atom system is strongly dependent on
the vibrational degrees of freedom of the system. 
As miniaturization decreases device sizes, the role of atomic vibrations
needs to be considered in the device functionalities.
This dependence has
been shown to profoundly alter the device behavior: from new channels
of conduction~\cite{Berthe} to heating  of
the junction~\cite{Neel}. It is then important to understand the role
of inelastic effects and the parameters that control them. 

A large
body of research has been devoted to inducing controled inelastic effects.
A recent review article~\cite{Ho2003} shows that inelastic effects in
a tunneling junction can be used to chemically analyze the molecules at the
junction by
inelastic electron tunneling spectroscopy (IETS)~\cite{Stipe},  
to induce  reactions by displacing atoms~\cite{Lee}, 
and to use molecular conformational changes as a switch for possible 
devices~\cite{Gaudioso}. In this way, the scanning tunneling
microscope (STM) induces the reaction
and also detects its product, a dramatic example is the modification
and detection of trans-2-butene on Pd (110)~\cite{Kim}.
The effect of local vibrations during electron
flow has also been
revealed in careful experiments of photon analyses~\cite{Hofluorescence}
yielding an unprecedented insight in vibrational dynamics.
Not only has the tunneling junction of an STM  been used to explore
the coupled electron-vibration dynamics, also point contact
spectroscopy has been used in molecular wires~\cite{Ruitenbeek,Agrait},
where the conductance showed drops at the molecular vibrational onset.

These many experimental results call for important theoretical development.
Indeed, the last years have seen several theoretical works spanning
most of the experimental systems: from tunneling inelastic
spectra~\cite{Mingo,Lorente,Ueba} to point contact 
spectroscopy~\cite{Todorov,Frederiksen,Solomon}, from systematic approaches
studying different parameters~\cite{Galperin2004,Galperin2006}
to lowest-order expansion with ab-initio parameters~\cite{Paulsson,Troisi}.
A thorough review on methodology and results can be found in Ref.~\onlinecite{Galperin2007}.

A quantitative description of inelastic processes is mandatory in order to
assess the relevance of the different ingredients characterizing
electron transport on the atomic scale. Recent developments in {\em ab-initio}
calculations  together with transport calculations permit us to grasp
the essential parameters and, eventually, produce predictive calculations.
Yet, the typical {\em ab-initio}-based calculations are on the one-hand-side 
heuristical, because a well established dynamical theory of
electron transport is yet to come~\cite{Prodan}, 
and on the other hand-side, they are complex and difficult to interpret.
Most {\em ab-initio} approaches use ground state density
functional theory with non-equilibrium Green's functions (NEGF), this
combination
is not justified, and the results have the full complexity of NEGF.
As signaled in Ref.~\onlinecite{Prodan},  new methodology
to treat quantum transport will be probably based in time dependent
density functional theory (TDDFT). Recent results show that it is
possible to treat nuclear (semiclassically) and electronic (quantally)
degrees of freedom within TDDFT to treat the non-adiabatic transport
problem~\cite{Almbladh,DiVentra}. New developments go
a step further in the treatment of correlated electron-ionic 
dynamics~\cite{McEniry}. 
But time-dependent methodology can have other
benefits beyond its correctness. In particular, it can be used to develop
a physical picture of the electron diffusion process, in this way
yielding complementary information to the more involved NEGF approach.
Time-dependent approaches can also have interesting numerical
behavior. Indeed, electron transport treated with the short-iterative
Lanczos method~\cite{SIL} has a quasi-linear scaling for sparse Hamiltonians.

In this article, we explore inelastic effects in electron transport
by means of electronic wave packet propagations in an idealized
atomic-size system. We consider tight-binding chains connected
to one and two vibrating electronic sites. These vibrating sites
can hold one and two nuclear modes that are coupled to linear order
in the nuclear displacement with the electronic degrees of freedom.
Despite the simplicity of this model system, the main one-electron
ingredients are included, and the time resolved solution permits us
to have insight different from the perturbation-theory Green's functions
results. Similar treatments have already been performed for
the case of electron-molecule collisions~\cite{Gauyacq} and
for inelastic effects in transport~\cite{Bringer}. 

The calculations presented here are distant from the experimental situation
because the model system is very simplified and because many-electron
problems are absent. Indeed, recent treatments show the richness of
effects associated with the many-electron aspects of the 
problem~\cite{Hyldgaard}, as well as the non-equilibrium many-phonon
problem~\cite{Flensberg,Mitra}. Despite, the absence of these very
interesting ingredients of inelastic transport, our calculations can
help in understanding inelastic effects because there are situations
in which one-electron transport is justified even in the
presence of inelastic effects~\cite{Imry}. To a certain degree,
our calculations are equivalent and complementary to those
of H. Ness~\cite{Ness2006} the main difference being that a time-dependent
approach is adopted in the present study.

\section{Time-dependent wave packet propagation}

Stationary electron transport does not need a time-dependent description. 
However,
insight on vibrational excitation processes
during the current flow can be gained by time-dependent calculations. 
Numerically,
a time-dependent description can benefit from the quasi-linear scaling 
using sparse
Hamiltonians. As in stationary descriptions, the bottleneck of the
calculation lies in the matrix time vector product of the Hamiltonian acting
on a system's vector. Hence, a time-dependent approach can yield
especial insight in the actual transport process in a realistic nanoscale
system  
by using a localized basis set that leads to a sparse Hamiltonian. This
strategy seems to be particularly appealing for the
description of conduction in nanowires~\cite{Troels} 
and nanotubes~\cite{Stephan}.
 In this section, we explore the general time-dependent methodology
using wave packets, beyond the Kubo linear theory of 
Refs.~\onlinecite{Troels,Stephan}.

A particularly interesting time-dependent approach to have information
on the full electronic trajectory is the short-iterative Lanczos (SIL) wave
packet propagation~\cite{SIL}.
The initial-value problem is solved by applying the evolution operator to the initial
 wave function, so that the solution reads, $ \Psi(t) = e^{-i \mathcal{H}t}
 \Psi(0)$, where $\mathcal{H}$ is the Hamiltonian of
the system under consideration (atomic units are used throughout,
$\hbar = m = e = 1$, unless otherwise specified). 
The use of this equation is inconvenient in the
case of large matrix dimensions because
it implies a diagonalisation. Instead, we prefer to use a numerical approximation
 which consists in successive infinitesimal evolutions of the wave function, 
$\Psi(t+\Delta t) = e^{-i \Delta t \mathcal{H}} \Psi(t)$.
In order to have an efficient implementation, the SIL method
truncates the Hilbert space to a subspace spanned by a few vectors.
The computation of this truncated subspace is the bottleneck of
the calculation because the matrix-time-vector product to generate the subspace
is performed on the total dimension of the problem. Despite the truncation,
the richness of the
full Hamiltonian spectrum is recovered by repeating this operation
for all the different time steps and propagating in this way the
initial wave packet.

Let $\Phi_{1}$ be a starting vector,
the Lanczos propagation method consists in the construction of a Lanczos matrix
following the recursion relation~\cite{Lanczos, Cullum},

\begin{equation}
 \beta_{j+1}\Phi_{j+1} = \mathcal{H}\Phi_{j} - \alpha_{j}\Phi_{j} - \beta_{j}\Phi_{j-1}, \quad j \geq 1
\end{equation}

\noindent where $\alpha_j$ are the new diagonal
 matrix elements of the new tridiagonal matrix and $\beta_{j}$ are
the new upper and lower diagonal 
matrix elements.
Since the recurrence relation
is started by $\Phi_{1}$, initial wave function, $\Psi(t=0)$,
then $ \beta_{1}=0$ and  $\Phi_{0}=0$. 
The vectors $\Phi_{j}$ are called Krylov vectors, and define the Krylov space of order $l$, by spanning the subspace given by $\{\Phi_j\}$, namely

 \begin{equation}
\mathcal{K}^l=Span\{\Phi_{1}, \mathcal{H}\Phi_{1},...,  \mathcal{H}^{l-1}\Phi_{1}\}.
\end{equation}

The Krylov vectors are orthogonal by construction, they may be normalized, and define 
thus a basis set in which the Hamiltonian can be expressed. In this basis, the 
Hamiltonian is tridiagonal. Additionally, the order $l$ can be much lower than the initial 
matrix dimensions. Consequently, the Lanczos matrix can be easily 
diagonalized using conventional algorithms. Regarding the Krylov vectors, it is 
known that for relatively large values of $l$, their orthogonality 
can be lost, this is the reason why we have set sufficiently small time steps  to keep a 
rather small order $l$ ($l \leq 9$).  
Its value can be changed at each step in order to optimize the 
performances of the calculation, lowering the computational cost  as
$l$ decreases. An 
excellent description of this dynamical control of the accuracy of the 
Lanczos propagation method can be found in Ref.~\onlinecite{Andrei}.

Another important feature is the undesirable effects provoked by the 
finite size of the propagation grid. To eliminate artificial reflections 
of the wave packet at the boundaries, 
we use a parabolic absorbing potential. 
%It is placed at both ends of
%the tight-binding grid representing both electrodes (please see next section
%for the model Hamiltonian used in this calculation)
%and started 400 sites before both edges. The potential is parabollic, 
%with a spring constant of $\frac{1}{400^2}$. 
In order to account for the errors which may be introduced by 
the reflections at the boundaries in the presence of the absorbing potential, 
we calculated the reflection coefficient with a broad wave packet. 
We found that less than 0.01\% of the wave packet was reflected at any energy.

To compute the energy-resolved transmission or reflection coefficients we use the
virtual detector technique~\cite{Detectors} which consists in the evaluation of 
the wave function some 
sites after the region where the electron-vibration interaction takes place. 
By a time-to-energy Fourier transform, 
transmission or reflection coefficients are obtained. 
In the inelastic case, this should be done for each vibrational state 
of the total wave function. Hence, for a 1-D tight-binding chain, 
the partial transmission, $\mathcal{T}_n(\omega)$, is given by

\begin{equation}
\mathcal{T}_n(\omega)=\frac{\mid \psi^{int}_{d,n}(\omega) \mid^2}{\mid \psi^{bare}_{d,0}(\omega)\mid^2}
\end{equation}

\noindent where we have assumed zero temperature and an initial wave packet in the
vibrational ground state, $n=0$.
\noindent $\psi^{int}_{d,n}(\omega)$ 
is the energy-resolved wave function in the vibrational state $n$, after the interacting region.
The detector has been located on site $d$. $\psi^{bare}_{d,0}(\omega)$ is the wave function computed with the same virtual detector but considering a bare Hamiltonian containing no vibrational degrees of freedom nor elastic defects.

To analyze quantities such as total transmissions, especially for non-trivial Hamiltonians which may have two or more vibrating electronic states, we have computed
 the density of states projected on any state, for instance, 
on any linear combination of the tight-binding states.
Consider a state $ \mid\alpha,  n\rangle$, with $\alpha$ a particular electronic state and $n$ its vibrational state,  evolving with Hamiltonian $\mathcal{H}$.
  The density of states projected on  $ \mid\alpha,  n\rangle$ reads~\cite{Tannor},

\begin{equation}
\rho(\omega)=\int_{-\infty}^{\infty}\frac{dt}{2 \pi} e^{i\omega t} \langle \alpha, n | e^{-i\mathcal{H}t} | \alpha, n  \rangle .
\label{PDOS}
\end{equation}

%%%%% Fin
 
%Sergio - Semana del 25/02/08

\section{The physical model}

The Fr\"ohlich-Holstein model~\cite{Holstein} 
is used to represent the combined electron-vibration
system: the electron-vibration coupling is assumed to be linear in the 
normal-mode coordinates. If we assume only one mode of vibration, and a single site, the Hamiltonian reads,

\begin{eqnarray}
\mathcal{H} = \varepsilon_0 c_0^{\dagger}c_0 + \sum_{k,i} \varepsilon_{k,i} c_{k,i}^{\dagger}c_{k,i} +  \sum_{k,i} t_{k,i}(c_{k,i}^{\dagger}c_{k+1,i}+ c_{k+1,i}^{\dagger}c_{k,i}) \\
\nonumber
 +\sum_{k,i} t_{k,i}(c_{k,i}^{\dagger}c_0 + c^{\dagger}_0 c_{k,i})
+\Omega b^\dagger b +  Mc^\dagger_0 c_0(b^\dagger + b)
\label{Hamiltonian}
\end{eqnarray} 

\noindent where $c^\dagger_k$ and $c_k$ are the fermionic operators which create and 
anihilate an electron in state $k$. 
Similarly, $b^\dagger$ and $b$ are the bosonic operators that create and anihilate a 
quantum of energy $\Omega$ in the considered vibrational mode. 
The first term in the Hamiltonian~(\ref{Hamiltonian}) refers to the site where the interaction 
takes place, it has an on-site 
energy of $\varepsilon_0$. The second describes the energy of site $k$ of chain $i$. 
For simplicity we will be dealing 
with only two chains ($i=L,R$: left and right). The third term describes the couplings 
among sites of chain $i$. Here, we just consider nearest neighbours. 
The fourth term is the coupling between the two semi-infinite chains 
and the state of energy $\varepsilon_0$.  The fifth term is the energy 
of the harmonic oscillator. Finally, the last term describes the electron-phonon 
interaction. In this single-state single-mode model, $M$ is only a scalar 
which represents the strength of the interaction. 
It is called the electron-phonon coupling. In the present work, we have explored 
both an electronic single site coupled to vibrations, and a double site. 
In the latter case, $M$ will be a $2\times 2$ matrix, as presented in section V.

Using a tensorial product description of the
electronic and nuclear coordinates, the full Hamiltonian
can be expressed in matrix form:
% Hamiltonian and basis set

\begin{equation}
\mathcal{H}=\left(
\begin{array}{cccccccccccccccccc}
 \mathcal{H}^{(0)} & \hat{M}                  & 0                            &                               &                             &                                  \\
 \hat{M}                &  \mathcal{H}^{(1)}  & \sqrt{2} \hat{M}      & 0                            &                             &                                  \\
 0                          & \sqrt{2} \hat{M}      &  \mathcal{H}^{(2)}  & \sqrt{3} \hat{M}      &  0                        &                                   \\
                             & 0                            &  \sqrt{3} \hat{M}     &  \mathcal{H}^{(3)} & \sqrt{4}\hat{M}     & 0                                \\
                             &                               &                               & \ddots                    & \ddots                  &\ddots                          \\
                             &                               &                               & 0                            &\sqrt{N-1} \hat{M} & \mathcal{H}^{(N-1)}    
 \end{array}
 \right)
\label{hinela}
 \end{equation}

\noindent where we have used a block representation in the vibrational basis $\lbrace | n\rangle \rbrace$.  The diagonal elements are electronic Hamiltonians, which define the propagation of a wave-packet in a vibrational subspace.
For clarity we give their tight-binding representation,

\begin{equation}
\mathcal{H}^{(n)}=\left(
\begin{array}{ccccccccc}
\ddots & \ddots & \ddots & &  & & &  &\\
0 & t & n\Omega & t & 0 & & & &  \\
 & 0& t &  n\Omega & T_l & 0 & & &  \\
 & & 0 & T_l & \varepsilon_0 +  n\Omega& T_r & 0 & & \\
 & & & 0 & T_r  &  n\Omega & t & 0 &  \\ 
 & & &  & 0 & t &  n\Omega & t & 0  \\ 
 & & & & & & \ddots & \ddots & \ddots
\end{array}
\right)
\label{HamHop}
\end{equation}

\noindent where $n$ labels a particular vibrational state in the harmonic approximation. 
The $t$ are off-diagonal matrix elements
which connect the nearest neighbours sites inside the left and right chain. 
The $T_l$ end $T_r$ play the same role as $t$, they connect
the chains to the site which has on-site energy $\varepsilon_0$. The diagonal term $n\Omega$ 
accounts for the energy the electron must exchange with the vibrational degrees of freedom 
of the system. If it propagates from one vibrational state to another, it must lose or gain 
$\Omega$,  the energy quantum of the vibration.

The matrices $\hat{M}$ in (\ref{hinela}) couple the Hamiltonians $\mathcal{H}^{(n)}$ 
in the different vibrational subspaces.
In the case of a single-site impurity,  $\hat{M}$ is
essentially a sparse matrix with the same dimensions as $ \mathcal{H}^{(n)}$, 
where only one single element is non-zero, the one connecting the diagonal 
matrix elements of energy $\varepsilon_0 +n\Omega$ and $\varepsilon_0+(n+1)\Omega$. 
In Hamiltonian~(\ref{hinela}), the factors that multiply $\hat{M}$ come from 
the matrix representation of the bosonic operators, they correspond
to the factors which appear in the well known relations 
$b^\dagger | n \rangle = \sqrt{n+1} | n+1\rangle$ and
$b | n \rangle = \sqrt{n} | n-1\rangle$. We note that  
Hamiltonian~(\ref{hinela}) is truncated, the number of vibrations that are considered 
in the calculation is $N$ in this example. This number $N$ may be sufficiently large to represent suitably the vibrational space. A detailed 
study of the convergency of the calculation is presented hereafter.

%Transmission & PDOS

Let us consider a molecule between two electrodes, modeled by a single state connected to two chains. In this single-site case, Y. Meir and N. S. Wingreen \cite{Meir} have shown that the current can be expressed as follows,

\begin{equation}
\mathcal{J}=-\frac{1}{\pi}\int \lbrack f_L(\omega)-f_R(\omega)\rbrack Im\lbrace tr\lbrack \Gamma G^r\rbrack \rbrace d\omega ,
\end{equation}

\noindent where $\Gamma$ is defined as a function of the couplings to the right and left leads,
 $\Gamma =\frac{\Gamma_R \Gamma_L}{\Gamma_R + \Gamma_L}$. $G^r$ is the retarded green function of the molecule, and $f_{L(R)}$
 is the Fermi distribution of the left (right) lead. 
This formula applies in the case where the couplings 
to the leads only differ by a constant factor, $\lambda$, such that 
 $\Gamma_L=\lambda \Gamma_R$. This is always so in the wide-band limit
for the single-site case. This is not the case
beyond the wide-band limit.
 
The equation above can be viewed as the integral of a quantity that we identify as a 
transmission, multiplied by an energy
window given by the difference of the Fermi distributions of the leads, meaning that a current will flow if both the transmission is non-zero
and the voltage applied between the electrodes is sufficiently large. Following the derivation by H. Ness \cite{Ness2006} at the zero temperature limit, the transmission reads,

\begin{equation}
Im\lbrace tr\lbrack \Gamma(\omega) G^r(\omega)\rbrack \rbrace = \Gamma(\omega)Im G^r_{00} ,
\label{traza}
\end{equation}

\noindent where $G^r_{00}$ is the projection of the Green function in the $n=0$ vibrational subspace, $G^r_{00}=\langle 0 | G^r | 0\rangle$. This means that the transmission is related to the projected density of states, Eq.~(\ref{PDOS}), in such a way that, if we consider the wide-band
approximation, where the coupling $\Gamma$ is independent of energy, the transmission is proportional to the density of states projected
in the $n=0$ subspace. By using the optical theorem one can explicitly retrieve the
vibrationally excited states in the inelastic current~\cite{Ness2006}.

In the case of the wave packet calculation, we can extend these equations for
the calculation of the electronic current. We approximate the
conductance assuming that the Fermi level only enters to define possible
final electronic states (otherwise the calculation is fully one-electron,
an in general we will not consider a Fermi level)

\begin{equation}
\sigma(\omega)=\frac{1}{\pi}\sum_n\mathcal{T}_n(\omega)\lbrack 
f_L (\omega+n \Omega) + f_R (n \Omega -\omega)\rbrack ,
\label{conductance}
\end{equation}

\noindent where the factor $\frac{1}{\pi}$ is the conductance quantum in atomic units. 
The terms between brackets are introduced
 to account for the opening of vibrational channels
since they contain the Fermi distribution
functions of the left electrode, $f_L$ and
the right one, $f_R$, at zero bias voltage in the present case. Actually, when electrons 
do not have sufficient energy to deposit it into the vibrational degrees of
freedom of the system, 
only the $\mathcal{T}_0(\omega)$ is to be considered in the conductance. 
The same holds if the energy of the incident electron is sufficient 
to excite one vibration, in this case we consider 
the sum of $\mathcal{T}_0(\omega)$ and $\mathcal{T}_1(\omega)$. 
Nevertheless, the calculation should be performed with a
sufficiently high value of $n$ in order to take into account 
the influence of closed channels in the conductance.
Finally, the current as a function of the voltage, $V$, can be written 
as the integral over energy of the conductance,

\begin{equation}
\mathcal{J}(V)=\int_{-\infty}^{\infty}\sigma(\omega)
\lbrack f_L(\omega) - f_R(\omega) \rbrack d\omega
.
\end{equation}

\noindent The terms between brackets ensures that the transmitted electrons go from occupied states to empty ones where the voltage dependence is included.
%%%%% Fin

% Comparison with Hervé Ness

The treatment presented here is complementary of the one by H. Ness~\cite{Ness2006}.
Indeed, the physical model is the same one, the difference is
the solution method. As in the case of Ref.~\onlinecite{Ness2006} 
the present approach is single-electron, neglecting
both electron occupation and electron-electron correlation effects.

\begin{figure}[tb]
\begin{center}
\includegraphics[width=0.95\columnwidth]{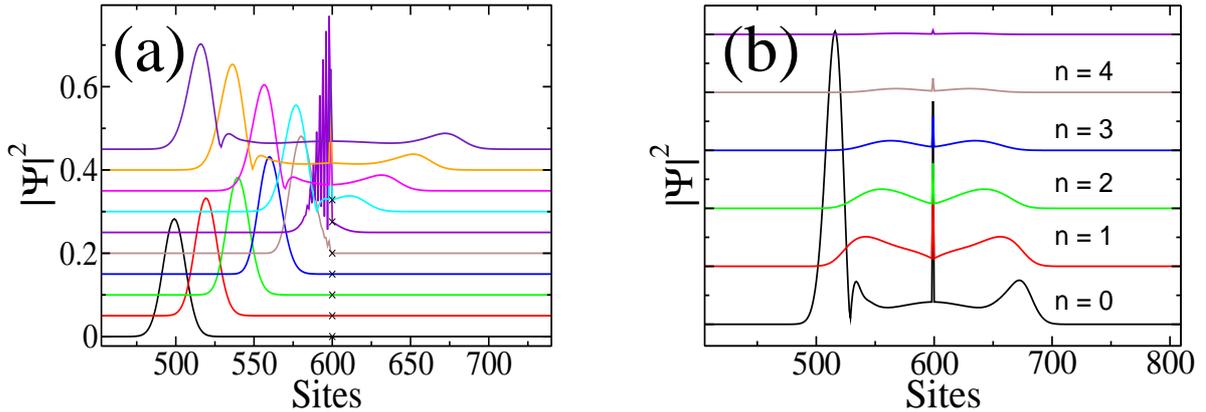}
\end{center}
\caption{(Color online)
(a) Spatial dependence for the squared modulus of the electronic wave function
at different times, arbitrarily shifted for representation
purposes. Case (a) is the elastic propagation of the wave packet, $n=0$,
where $n$ is the vibrational quantum number. Symmetric hopping
matrix elements, $T_L = T_R$ have been used.  
In (b) the wave packet is represented at the same instant in time
for different vibrational states $n$.
The wave packet is initially propagated from the left electrode in the
vibrational ground state, $n=0$. The calculations are
performed at zero temperature. When the wave packet starts populating
the site number 600 (marked with a cross in (a)), the different
vibrational states are populated (b), and the vibrationally excited
wave packet starts propagating in both directions, indistinctly, due
to the symmetric electronic couplings. 
}\label{Packet}
\end{figure}

\section{Single-site resonant model results}

%Typical case

Let us assume a single state coupled to the two
1-D tight-binding chains of the above model.  
Figure~\ref{Packet} shows the results of a wave packet propagation.
Figure~\ref{Packet} (a) is the elastic wave packet, $n = 0$, because
we assume that no vibration is initially excited in
the system and the temperature is zero. The present system is two 1-D
electodes symmetrically coupled to the vibrating site. Hence, 
in Fig.~\ref{Packet} (b) we see that only the reflected wave packet
is different from the transmitted one in the elastic
channel, $n = 0$, while it is identical for the inelastic channels:
the electron reaches the active site and populates the ladder of
excited vibrational states. In each state the electron has a finite
and identical left and right transmission.

\begin{figure}[tb]
\begin{center}
\includegraphics[width=0.95\columnwidth]{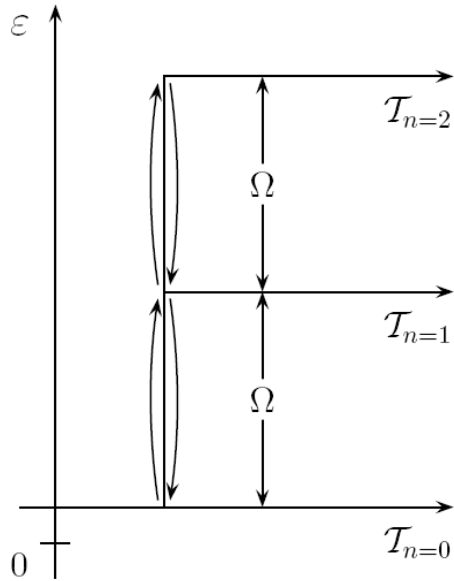}
%\pnode(3,-6){A}
%\pnode(9,-6){B}
%\ncline[arrowsize=5pt]{->}{A}{B}
%\pnode(2,-3){C}
%\pnode(5,-3){D}
%\pnode(5.1,-2.9){P}
%\pnode(5.1,-3.1){Q}
%\pnode(9,-2.9){R}
%\pnode(9,-3.1){S}
%%\ncline[linestyle=dotted]{->}{P}{R}
%%\ncline{->}{Q}{S}
%%\pnode(6,-5.9){E}
%\pnode(5,-6){E}
%\pnode(5.1,-5.8){L}
%\pnode(9,-5.8){M}
%%\ncline[linestyle=dotted]{->}{L}{M}
%\ncline{-}{D}{E}
%\pnode(9,-5.9){F}
%%\ncline[linestyle=dashed]{->}{E}{F}
%%\pnode(2,0){G}
%\pnode(5,0){H}
%\pnode(5.1,-0.2){N}
%\pnode(9,-0.2){O}
%%\ncline{->}{N}{O}
%%\ncline[linestyle=dotted]{-}{G}{H}
%\pnode(9,0){I}
%\ncline[arrowsize=5pt]{->}{H}{I}
%\pnode(9,-3){J}
%\ncline[arrowsize=5pt]{->}{D}{J}
%\pnode(5,-0.1){K}
%\ncline{D}{H}
%\pnode(9,-0.1){L}
%%\ncline{->}{K}{L}
%\ncarc[nodesep=3pt, offset=2pt, arrowsize=5pt]{->}{E}{D}
%\ncarc[nodesep=3pt, offset=2pt, arrowsize=5pt]{->}{D}{E}
%\ncarc[nodesep=3pt, offset=2pt, arrowsize=5pt]{->}{D}{H}
%\ncarc[nodesep=3pt, offset=2pt, arrowsize=5pt]{->}{H}{D}
%%\pnode(2,-3){T}
%%\ncline[linestyle=dashed]{T}{D}
%\pnode(3.5,-7){U}
%\pnode(3.5,1){V}
%\ncline[arrowsize=5pt]{->}{U}{V}
%\rput(3,0.7){\Large{$\varepsilon$}}
%\rput(8.5,-0.5){\large{$\mathcal{T}_{n=2}$}}
%\rput(8.5,-3.5){\large{$\mathcal{T}_{n=1}$}}
%\rput(8.5,-6.5){\large{$\mathcal{T}_{n=0}$}}
%\pcline[offset=4pt, arrowsize=5pt]{<->}(6.5,-6)(6.5,-3)
%\ncput*{\large{$\Omega$}}
%\pcline[offset=4pt, arrowsize=5pt]{<->}(6.5,-3)(6.5,0)
%\ncput*{\large{$\Omega$}}
%\pnode(3.3,-6.5){W}
%\pnode(3.7,-6.5){X}
%\ncline{W}{X}
%\rput(3,-6.8){\large{$0$}}

\end{center}
\caption{ Vibrational energy level scheme: The electronic wave packet
propagates from left to right in the $n=0$ vibrational ground state.
When the electron wave packet populates the impurity site connected to
the vibrator, the population of $n=1, n=2, ...$, becomes different
from zero and the wave packet propagates in both directions. The transmitted
wave packet permits us to compute the transmission resolved in $n$.
The transmitted wave packet can be assigned to a vibrational channel. The
$n=0$ channel is the elastic one, that can have contributions from all
channels due to the excitation and de-excitation of the vibration during
the wave packet propagation. This leads to a rich vibronic structure in the
electron transmission even for the elastic channel.
}\label{schema1}
\end{figure}

% Meaning of the transmission function
A simple-minded picture of the structure appearing in the transmission
as a function of incident electron energy
 could be that  there is a series of N resonances displaced
by the phonon energy, in agreement with the scheme of Fig.~\ref{schema1}.
 At a given
inital electron energy, the wave packet probes the resonant electron
site if the energy is within $n\Omega$ of the resonant site energy.
The results is a series of equidistant peaks spaced by $\Omega$.
This is plotted in Fig.~\ref{Transmission} (a). There, the transmission for
the single electronic state without electron-vibration coupling
is depicted in a dotted line, and the full line corresponds
to the case when the electron-vibration coupling is included.
 
However, the physical picture is more complex than just a series
of equidistant resonances.
In order to understand the appearance of the vibrational sidebands
we divide the transmission according to the final state of the vibrator,
Fig.~\ref{Transmission} (b).
This yields information on the vibrational state once the wave packet
has propagated through the resonant site. The result is that each
transmission curve for a well-defined final vibrator state displays
a similarly rich peak structure. The above picture, Fig.~\ref{schema1}, has to be changed:
there are complex vibrational pathways, in which different
vibrational states are probed before the system is left in
a singly well-defined vibrational state. Given the coherence of
the electron propagation, the different pathways can interfere
and will give rise to Fano lineshapes in the transmission function~\cite{Durand}.

\subsection{Vibrational state population sequence: electron coherence}

\begin{figure}[tb]
\begin{center}
\includegraphics[width=0.95\columnwidth]{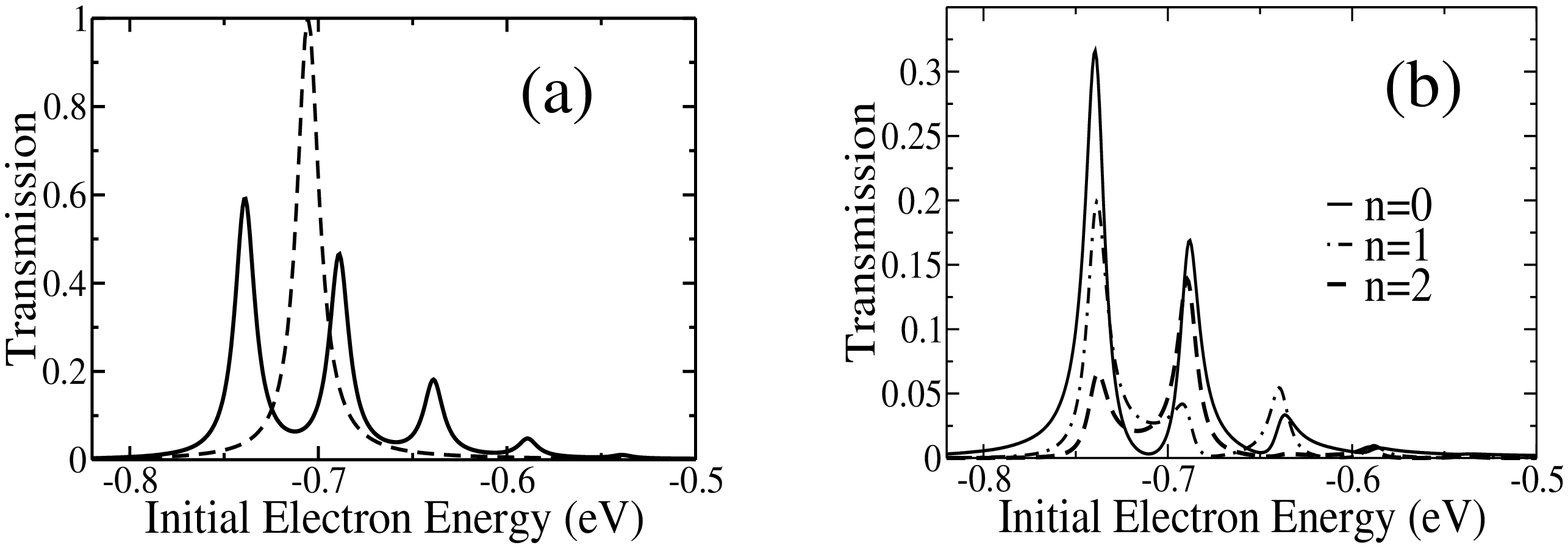}
\end{center}
\caption{ (a) Electron transmission for two 1-D tight-binding electrodes
connected by a single electronic site with electron-vibration coupling 
(full line) and in its absence (dashed line). (b) Tranmission decomposition
in the final vibrational state of the single site, 
$n$ is the number of vibrational quanta.
The elastic transmission is the $n=0$ curve in (b).
The initial state $\varepsilon_0 = -0.7$ eV, $\hbar \Omega = 0.05$ eV,
the electron-vibration coupling is $M = 0.04$ eV and the resonance
width is set by the hopping matrix elements $T_l = T_r = 0.05$ eV. The
bandwidth is 2 eV.
}\label{Transmission}
\end{figure}

Wavepacket dynamics can probe the different structures appearing 
in the transmission
function in order to yield information on the actual interference process.
Let us take a vibrational ground state,
 broad electronic wave packet energetically centered about the first maximum of the
full transmission function, Fig.~\ref{Transmission}. 
A spatially broad wave packet is almost a planewave and synonymous 
of a monochromatic packet, hence we can be sure to probe only the structure
of the first maximum. 


Figure~\ref{Population}  shows the modulus square of the electronic wave function
for the first three vibrational levels on the vibrating site as a
function of time. We first notice that the population
of the vibrations are sequential: the peaks are slightly shifted in time as
$n$ increases. The linear electron-vibration coupling forces this sequential
population since $n$ is changed in steps of one quantum of vibration, Eq. (\ref{Hamiltonian}).
The populations show a time modulation, in the present case, we see that the modulation
frequency of the $n=1$ wave packet is roughly twice the modulation 
of the ground state and of $n=2$. This is easily explained by considering populating
and depopulating the $n=1$ level by populating the $n=0$ and the $n=2$. After a certain
time the $n=2$ level can depopulate again in the $n=1$ as well as the $n=0$. Despite
the simplicity of the electron-vibration coupling and the electronic model, the final
population dynamics depend on the wave packet energy, the strength of the vibrational
coupling and the electron lifetime at the site, leading to non-trivial dynamics.
The lifetime of the site is fixed by the hopping matrix elements $T_l$ and $T_r$, Eq. (\ref{HamHop}).

%Rate equation approach

The Fano profile characterizing the transmission functions, Fig.~\ref{Transmission}, can then
be explained by the interference of several of these vibrational pathways. Indeed,
asymmetric Fano profiles are the rule in Fig.~\ref{Transmission} (b). 
For the case of the total transmission,
the addition over final vibrational states smear out the different profiles, rendering
more difficult the determination of an asymmetric Fano profile in the peak sequence.

\subsection{Time delay and phase shifts}

Spatially narrow wave packets contain most of the energy components
needed to span the spectral region of interest, i. e. the main peaks of 
Fig.~\ref{Transmission}.
The Fourier coefficients in the energy domain can be evaluated to obtain the
phase of each component and hence the phase shift between Fourier components.
The derivative of the phase shift with respect to energy yields the
time delay of the Fourier component in the wave packet~\cite{Gauyacq}. 
Hence, we can analyze
the effect of the vibrational excitation of the electronic site
by means of the time delay imposed on the propagating wave packet.

\begin{figure}[tb]
\begin{center}
\includegraphics[width=0.95\columnwidth]{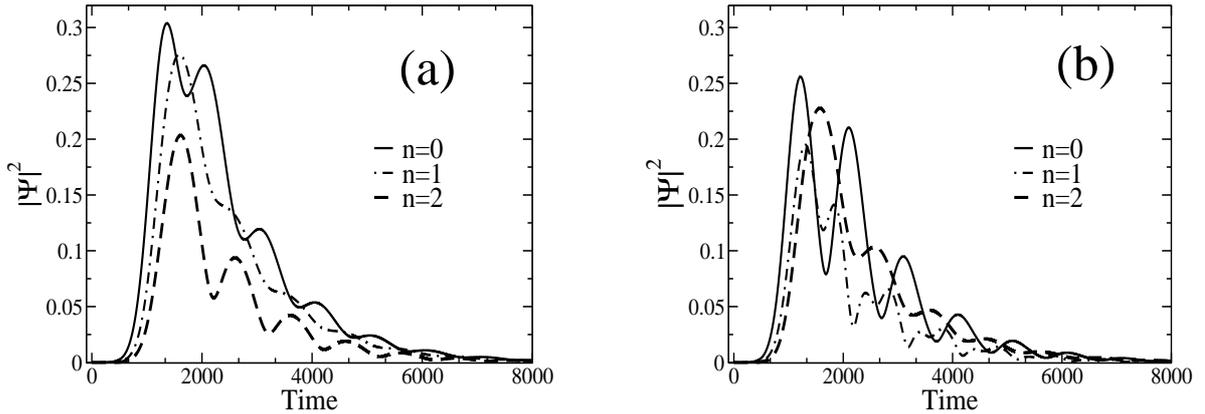}
\end{center}
\caption{Time dependence for the squared modulus of the electronic wave function
at the site connected to the vibrator. In case (a) an almost
monochromatic wave packet has been centered at the energy of the lowest peak
of Fig.~\ref{Transmission} (b). In the case (b) the wave packet has been
centered at the second peak, where the transmission in $n=1$ is
smaller than the transmission in $n=2$.
}\label{Population}
\end{figure}

\begin{figure}[tb]
\begin{center}
\includegraphics[width=0.95\columnwidth]{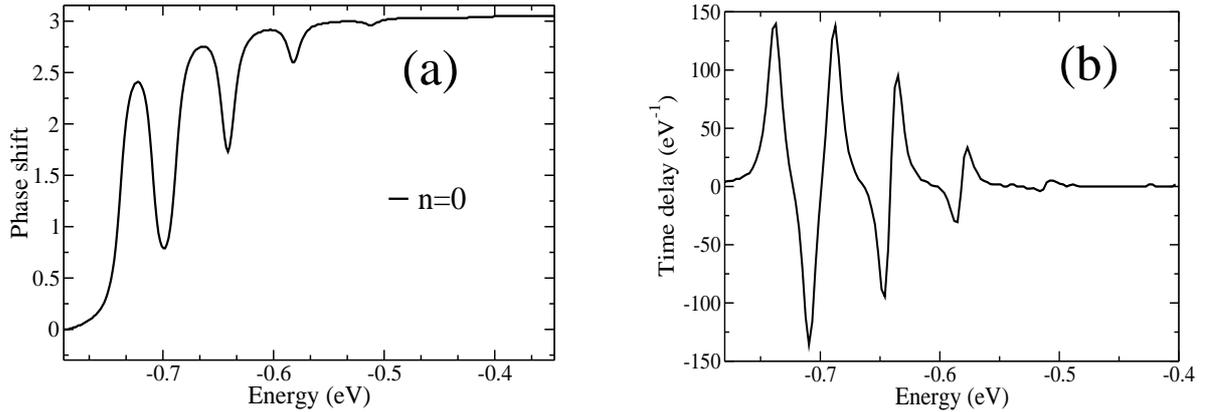}
\end{center}
\caption{
Electronic phase shift (a) and time delay (b) for the elastic channel, $n=0$,
for the case of Fig.~\ref{Transmission} (b). The phase shift over the full
resonant structure amounts to $\pi$, but the vibrational substructure leads
to fast variations of the phase, showing the resonant nature
of the vibronic peaks of Fig.~\ref{Transmission}. The time delay is maximum away
from each resonance, but it can change signs on resonance, depending on the order of the
vibronic peak. This delay/acceleration process is related to the interference
path inside the vibrator.
}\label{PhaseShifts}
\end{figure}

Figure~\ref{PhaseShifts}(a) shows the phase shift for the elastic component of the wave packet (the $n=0$
component). The full phase shift over all of the vibronic peaks is $\pi$, this
is the phase shift of the electronic resonance. The phase shift of each individual
contribution to the transmission is more complex. We see that the phase shift varies
rapidly after each maximum,  very much indicating a suite of Breit-Wigner-like resonances. 
Indeed, the the phase shift $\delta$ for a Breit-Wigner resonance
of FWHM $\Gamma$ centered at $E_0$ is given by~\cite{Gauyacq}:
\[\delta = - tan^{-1}\left (\frac{\Gamma}{2(E-E_0)} \right ). \]

The time delay, $\tau$, is given by the energy derivative~\cite{Gauyacq}
 of the phase $\delta$:
\[\tau=\frac{d\delta}{dE}. \]
In the case of a single Breit-Wigner resonance, the time delay is then
\[\tau=\frac{\Gamma/2}{(E-E_0)^2+(\Gamma/2)^2}.\]
On resonance the time delay is just $\tau=2/\Gamma$, yielding direct information
on the resonance width, $\Gamma$.
In the present case, the time delay, Fig.~\ref{PhaseShifts}(b),
 gives some definite interpretation: the
time delay is maximum once the wave packet has encompassed one of the vibronic
resonances. Hence, just above the resonance, the electron is deterred
in its propagation by interaction with the vibration. The vibration contributes
to the partial width of the electronic resonance, however the total
width is independent of the vibration~\cite{Wingreen}. A spatially narrow wave packet,
will have the phase shift and time delay of an electronic resonance regardless
of the existence or not of the vibrations. 
We also find negative time-delays, that correspond to drops in the phase shift. 
Electrons can hence be expelled from the resonance more easily in the presence
of vibrations at certain incident energies.
This behavior is due to the interference between two consecutive vibronic
resonances. As emphasized in Ref.~\onlinecite{Gauyacq}, the time delay has
to be interpreted with care in the case of a multichannel problem such as the present one.
The time dependence is indeed complex and the time delay is just a number
that cannot summarized the full intereference pattern.

The interferences among vibrational paths are ubiquituous in all present results.
Their effects can be
seen in the Fano-like lineshapes of the tranmission function Fig.~\ref{Transmission},
in the oscillating population of vibrational states with time, Fig.~\ref{Population} and
in the rapid changes of the phase shifts and in the negative time delays, Fig.~\ref{PhaseShifts}.

\subsection{Time scales}

To develop an understanding of the vibrational excitation process, many authors
resort to comparing the different timescales in play
\cite{Galperin2008,Galperin2007,Baratoff}.
This is a very appealing approach because it permits us to rationalize
the excitation process and the excitation regime of different
systems.

Let us define $\tau_{\mbox{mol}}$ as the lifetime of an electron in the molecule,
which is given by  the inverse of the resonance width, $1/\Gamma$.
Typically, for chemisorbed molecules, $\Gamma \sim 1-3$ eV, which means
electron lifetimes in the sub-femtosecond range.
In order to estimate the change in conductance due to an inelastic
process, we can compute the inelastic fraction of electrons
passing by the molecule. This can be estimated by computing the ratio
of the excitation time to the
first quantum of vibration, $\tau_{\mbox{n=1}}$, and the lifetime of
the electron in the molecule, $\tau_{\mbox{mol}}$.
Let us estimate the magnitud of the electron-vibration coupling, $M$, to attain
a measurable change in conductance (larger than 1\%).
This is the regime where perturbation theory
is valid. In this case, the excitation
process has been classified as sudden in the electron-molecule collisions literature~\cite{GauyacqLibro}.

In a sudden process, the molecule is assumed to be very briefly in its negative
ion potential energy surface (PES). This can be indeed very brief as has been said
above. 
As in Ref.~\onlinecite{Berthe}, 
let us assume that the PES
is basically parabolic, but displaced with respect to the neutral PES~\cite{Negative}:
\begin{equation}
E_- = \frac{1}{2} K (Q-Q_-)^2
\label{NegPES}
\end{equation}
where $K$ is the spring constant of the PES related to the mode frequency by
$ K = \mu \Omega^2 $ 
where $\mu$ is some reduced mass (this is easily generalized
to the case of multinuclear modes). Here, $Q_-$ is the displaced center of the
negative PES. When the electron flows through the molecule, the molecule is suddenly
in its negative PES. Hence, the nuclei experience a force given by
\[
F = -  \frac{\partial E_-}{\partial Q}|_{Q=0}. 
\] 
Then, they acquire a speed, $v$, of the order of
\[
v \approx - \frac{1}{\mu} \frac{\partial E_-}{\partial Q}|_{Q=0} \times  \tau_{\mbox{mol}}
\]
and $\tau_{\mbox{mol}} \approx 1/\Gamma $. Hence,
\[
v \approx \frac{ K Q_-}{\mu \Gamma},
\]
where $K Q_-$ is the electron-vibration linear coupling, $M/\delta Q_{rms}$ of Eq.~(\ref{Hamiltonian})
with $\delta Q_{rms} = 1 / \sqrt{2 \mu \Omega}$ coming from the second
quantization of the displacement $Q$.
The speed gained by the nuclei at an excitation of one quantum of vibration  near ${Q=0}$
is $ v_M = \sqrt{ 2 \Omega / \mu}$. In this way, we find an upper limit for $\Gamma$:
\begin{equation}
\Gamma \leq |M|.
\label{condition}
\end{equation}
This simple estimation shows that for strong electron-vibration coupling, vibrational
excitation is unavoidable, where strong means larger than the molecule-electrode
coupling. 

Weak coupling is then the regime when $|M| \ll \Gamma $. We can use Fermi's golden rule to
estimate the vibrational excitation rate~\cite{Mats}:
\[
\frac{1}{\tau_{{n=1}}} \approx 4 \frac{|M|^2}{\Gamma}.
\]
Assuming typical IETS branching ratios  
$\tau_{\mbox{mol}}/\tau_{\mbox{n=1}}\approx 10^{-2}$, 
leads to 
\[
\frac{|M|^2}{\Gamma^2} \leq \frac{1}{400} ,
\]
therefore the coupling becomes $|M| \sim \Gamma/200$. 
For chemisorbed molecules this is in the range of 0.01 eV. 
This coupling is easily found
in a large class of molecules, and therefore IETS is a feasible spectroscopy.  This
is a simple estimation showing that rather small $M$ ($\approx 0.01$ eV)
can have a measurable change in conductance for molecules adsorbed on metallic leads.
\\

One more time scale is fixed by the vibrational frequency, $\Omega$. As discussed in Ref.~\onlinecite{Gauyacq},
there are three basic regimes that can be distinguished:
\begin{enumerate}
\item $\Gamma \ll \Omega$ the negative ion is long lived and the molecule can vibrate in the new state,
\item $\Gamma \gg \Omega$ the negative ion is short lived and we are in the above sudden regime,
\item $\Gamma \approx \Omega$ strong interference effects appear due to the nuclear motion.
\end{enumerate}
The time scale
given by $\Omega$,
 defines then the type of electron-vibration interaction that will result. In the first
case, the electron has ample time to interact with all the nuclear degrees of freedom and 
depending
on the electron-vibration coupling, $M$, vibronic signatures can appear. 
This has been shown in the
case of STM studies of electronic states on surfaces~\cite{Berthe,Repp} with a
band gap leading to small $\Gamma$. 
The second case corresponds to the IETS case discussed above.
The third case has been studied in the gas phase (see for example Ref.~\onlinecite{Gauyacq}) 
but we are not
aware of any report on the consequent {\em boomerang} effect in the electron transport regime.

\subsection{Finite-band effects}

The wide-band approximation simplifies the expression
for the transmission  that becomes analytical~\cite{Wingreen}.
One of the results of the wide-band approximation is that the transmission
function and the spectral function or PDOS, Eq.~(\ref{PDOS}), have the same behavior with the electron
energy. This is definetly not the case when finite-band effects are considered,
indeed, the energy dependence of the coupling to the electrodes leads the
transmission function to have a different energy behavior than the site
spectral function, Eq.~(\ref{traza}). 
When the initial level, $\varepsilon_0$ is near the center
of the band, the conditions for the wide-band limit are easily attained. As
the level approaches one of the band edges, $\Gamma$ becomes strongly
dependent of energy as well as the center of the transmission distribution,
Fig.~\ref{Transmission}.

\begin{figure}[tb]
\begin{center}
\includegraphics[width=0.75\columnwidth]{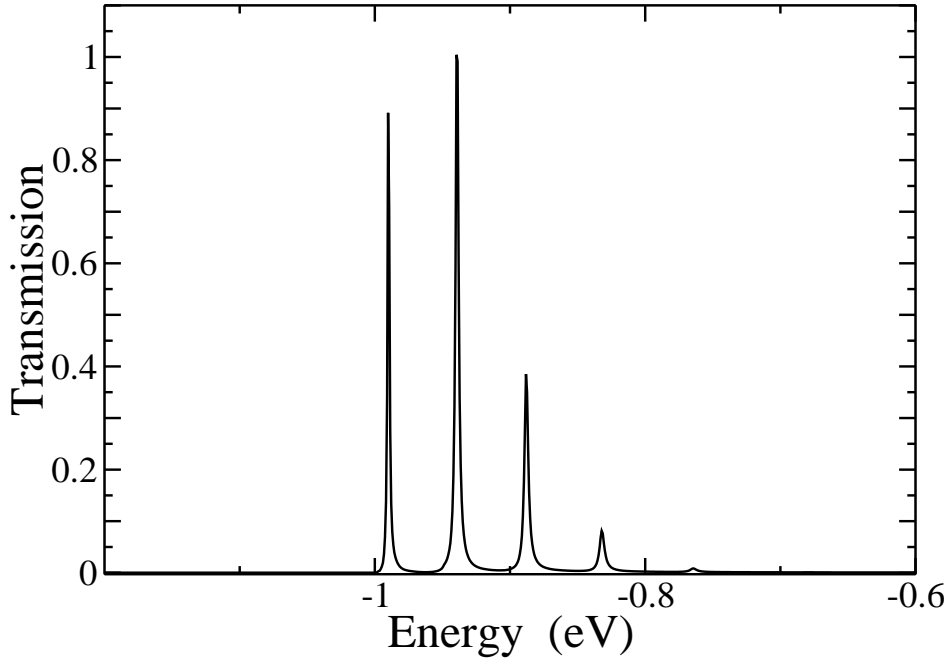}
\end{center}
\caption{
Transmission function for the same system of Fig.~\ref{Transmission}
where $\varepsilon_0$ is -0.95 eV instead of -0.7 eV. The closing of
channels due to the band edge leads to sharper and singular-like transmission peaks.
The band edge is at -1.0 eV.
}\label{fondo}
\end{figure}

One of the apparent features is that the transmission peaks become thinner
and better defined. As $\varepsilon_0$ approaches the band edge, many inelastic
channels
start closing because the final electron energy
falls out of the electron band, leading to a substantial decrease in the transmission peaks.
This is clearly seen in Fig.~\ref{fondo}. The parameters of the transmission
calculated in Fig.~\ref{fondo} are exactly the same as for Fig.~\ref{Transmission} (a)
except that $\varepsilon_0$ is -0.95 eV instead of -0.7 eV. The bottom of the
band is at -1 eV and the quantum of vibration is $\hbar \Omega = 0.05$ eV. The
singular aspect of the peaks is enhanced by the 1-D 
character of the electrodes. Indeed, when the final energy of the electron approaches
the bottom of the band a van Hove singularity appears in the density of states
making the transmission more singular.
In more realistic cases with 3-D electrodes we expect less singular transmission
peaks.

For an electron resonance approaching the top of the band, the situation
is rather different. Here, the higher vibronic side bands dissapear but
the $n$ convergency remains the same because
no channel is closed, leading to low-energy side bands of the
same width as for the case of the resonance at the center of the band.

\subsection{The dynamical polaron shift}

The polaron shift is defined as the energy displacement of the first peak in 
Fig.~\ref{Transmission} (a) with respect to the dashed-line peak. The polaron shift
is related to the appearance of a vibronic structure. As
we showed above, in the time-scale discussion, 2 conditions need be satisfied.
Namely, $M$ must be sizeable and $\Gamma \ll \Omega$. This leads to considering~\cite{Hyldgaard}
 the
parameter $p=M^2/(\Gamma \Omega)$. When $p >1 $ the polaron shift becomes measurable.

Hyldgaard et al.~\cite{Hyldgaard} show that the polaron shift critically depends on the initial
occupation of the vibrating electron site. When the electron site is occupied, the transmission
function shows a main peak displaced by $-3 M^2/\Omega$. At half-filling the displacement is 
$-2 M^2/\Omega$ and
for an empty site the polaron shift is $- M^2/\Omega$. 

We can give a direct interpretation of the polaron shift by using the above parabolic model, 
Eq.~(\ref{NegPES}). Following the above discussion, the polaron shift for the empty site is
$- M^2/\Omega$ which is equal to $-\frac{1}{2} K Q_-^2$. The neutral PES is given
by $\frac{1}{2} K Q^2$, hence the polaron shift is exactly the energy difference between the
neutral PES and the negative one at the nuclear coordinate at which electron
capture takes place. Hence, the polaron shift is the energy gain in the formation
of the negative intermediate because the electron-vibration coupling permits
the electron to probe the negative ion PES. In the absence of coupling, the nuclear
coordinates do not evolve, and the negative ion resonance is just a simple lorentzian.

In order to probe the polaron peak, the nuclear wavefunctions in the two ground states (neutral
and negative) need to overlap. The overlap is large at weak electron-vibration coupling, hence
a well-defined peak appears, just shifted by the energy gain. As the coupling increases, the
overlap diminishes, leading to a decrease of the polaron peak.
In the limit
of strong coupling, the polaron peak is basically zero and a rich structure of evenly distributed
peaks of the negative-ion vibrational states is apparent, with a gaussian envelop~\cite{Repp,Mahan}.

\begin{figure}[tb]
\begin{center}
\includegraphics[width=0.85\columnwidth]{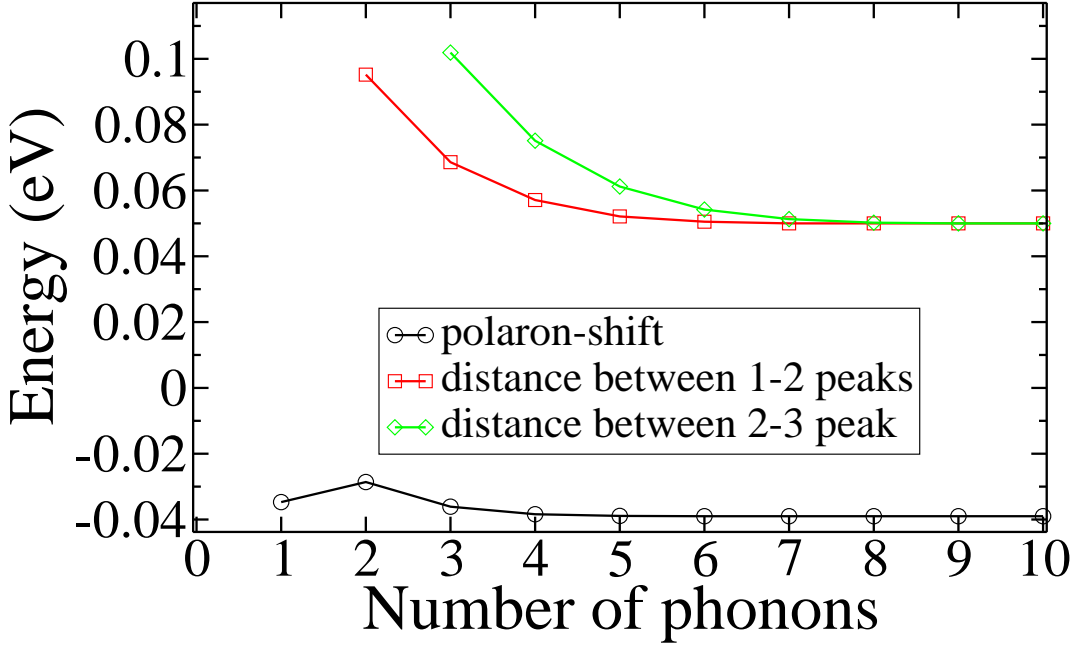}
\end{center}
\caption{(Color online) Convergency in the polaron shift and inter-peak distance
with the number of vibrational states included in the calculation.
The polaron shift is the displacement of the first peak with respect to
the resonance without electron-vibration coupling, 
dashed lines in Fig.~\ref{Transmission} (a). It is a critical measurement 
of the strength of the electron-vibration coupling, and hence, 
sensitive to convergency. Also sensitive is the inter-peak
distance that is equal to $\Omega$.
}\label{conv}
\end{figure}

% Figure on convergency with number of n
The polaron shift is a good test for the convergency of 
numerical calculations with the number of vibrational states 
included in the expansion of Eq.~(\ref{hinela}). 
Figure~\ref{conv} 
shows the convergency of the polaron shift with the number N of phonons included in the calculation. 
A correct value is obtained ($\approx -0.038 eV$) at about N=5 phonons. Following the results of the 
calculation of Hyldgaard et al. for an empty state, the polaron shift would be $-0.032 eV$ with our 
parameters ($M=0.04 eV, \Omega=0.05 eV$)

We also show in Fig.~\ref{conv} the inter-peak distance as a relevant quantity regarding the convergency of the calculation.
We observe that the peaks of lower energy tend to converge with less phonons than the peaks of higher energy, which need 
a higher number of phonons to be placed at $\Omega$ the one with respect to the 
other. Trivially, higher-order peaks need higher $n$, otherwise the peak may not even 
appear in the calculation.

%Figure with convergency regarding the site energy/number of Lanczos states
%Similarly, the convergency of the Lanczos algorithm is also revealed by studying the polaron shift. Figure~\ref{conv}

%%%%% Fin

The polaron shift is also very sensitive to the 
different approximations that are used when evaluating
inelastic effects in electron transport. Indeed, 
the self-consistent Born approximation (SCBA)~\cite{Hyldgaard} 
yields wrong polaron shifts.
Such a study is perfomed in~\onlinecite{Ness2006}. There, it is shown that
the SCBA is equivalent to neglecting the $\sqrt{n}$ factors of Eq.~(\ref{hinela}),
this leads to wrong interpeak distances. Nevertheless, the SCBA captures
much of the physics of inelastic processes and is an all-order theory,
becoming reliable enough and very interesting for the evaluation of
inelastic processes in a wide range of problems.
In the particular case of weak electron-vibration
coupling, $n \approx 1$ and the SCBA is virtually exact.

\section{Two-site resonant model results}

Most of the literature devoted to transport in the presence of electron-vibration
coupling is based on the single-site model. 
However, Bringer et al.~\cite{Bringer} showed that
the single-site case has a behavior that cannot be extended
to more realistic systems in which several electronic states 
are coupled with vibrations.
Here, we will analyze when one can reduce the problem 
to the single-site case and when
not. In the same way, we will show that when several vibrations are involved, the
neglection of some of the vibrations can lead to qualitatively wrong results.

\subsection{Two-sites and one vibration}

Our previous one-site model is extended to have two electronic sites connected
to a vibration. This model has been recently explored
in the context of ab-initio based calculations of inelastic
transport between two pyramidal electrodes~\cite{Frederiksen2007}. The
simplified two-site model permits us to understand the more complex
ab-initio results. The model is given again by  Hamiltonian~(\ref{Hamiltonian})
and (\ref{hinela})
where $\varepsilon_0$ and $M$ are $2 \times 2$ matrices.

Here, one only vibration is assumed to interact with the electron
flow. In the case of weak coupling~\cite{Frederiksen2007}, this 
assumption is justified because Eq.~(\ref{Hamiltonian}) is truncated
to the first phonon of each mode and the inelastic effects of
each mode are separable. In the next section, we will see that when
convergency in the number of phonons needs more than one
excited vibration, all of the vibrations need to be considered at once.

\begin{figure}[tb]
\begin{center}
\includegraphics[width=0.75\columnwidth]{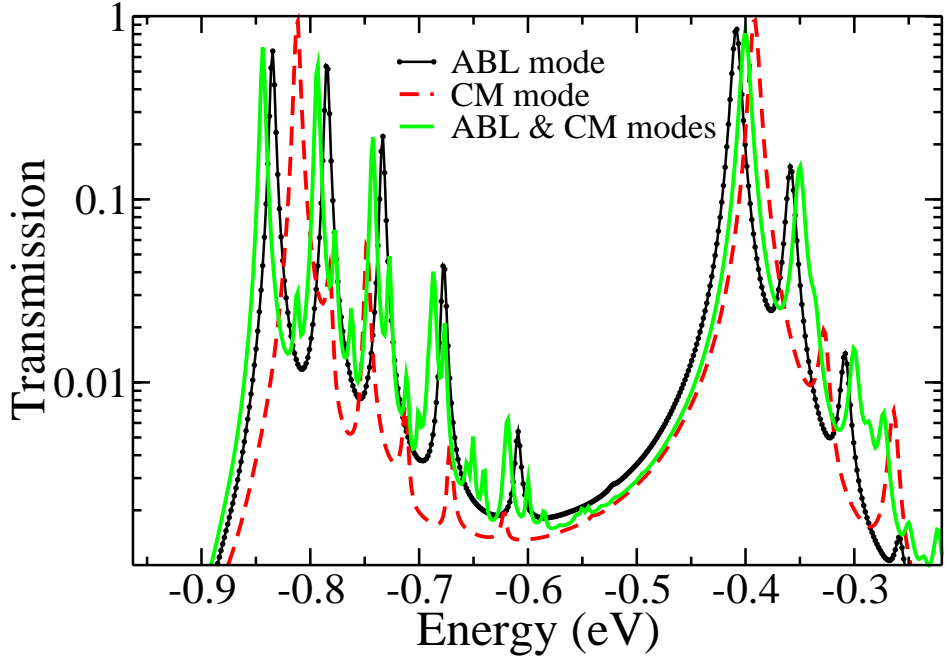}
\end{center}
\caption{
(Color online)
Electron transmission as a function of incident energy for the two-site
problem in semilog scale. Two different modes are consider, the anti-bond-length
modes (ABL) in full line with dots, and the center-of-mass mode (CM) in
dashed lines. These two calculations are performed independently, however
when both modes are simultaneously considered the transmission function is different
and becomes the full line marked by ABL \& CM modes. We see that no direct assignments
of the transmission peaks can be performed and the vibronic structure grows in
complexity beyond a superposition of transmission peaks from both modes.
}\label{ABL-CM}
\end{figure}

In this one-dimensional model, there are two only possible 
modes~\cite{Frederiksen2007}.
We will call them the center-of-mass mode (CM),
 and the anti-bond-length mode (ABL). The
CM mode means that both sites displace in phase,
 hence the electron-vibration coupling matrix have non-zero
 onsite elements (the diagonal):
\begin{equation} 
{\bf M}_{CM}=\left[ \begin{array}{cc}
m_3 & 0 \\
0 & -m_3 \end{array} \right],
\label{CM}
\end{equation} 
and the second element is negative in order to account for the sign of the
displacement of each site.
In the ABL mode, the sites are moving out of phase, corresponding to 
an internal stretch mode. Hence, the electron-vibration matrix becomes
\begin{equation} 
{\bf M}_{ABL}=\left[ \begin{array}{cc}
m_1 & m_2 \\
m_2 & m_1 \end{array} \right].
\label{ABL}
\end{equation} 

Figure~\ref{ABL-CM}  shows the transmission for the 
ABL and
CM modes, independently computed and together (to be analyzed in the
next section).
These figures show  series of vibronic peaks
difficult to analyze. 

\begin{figure}[tb]
\begin{center}
\includegraphics[width=0.95\columnwidth]{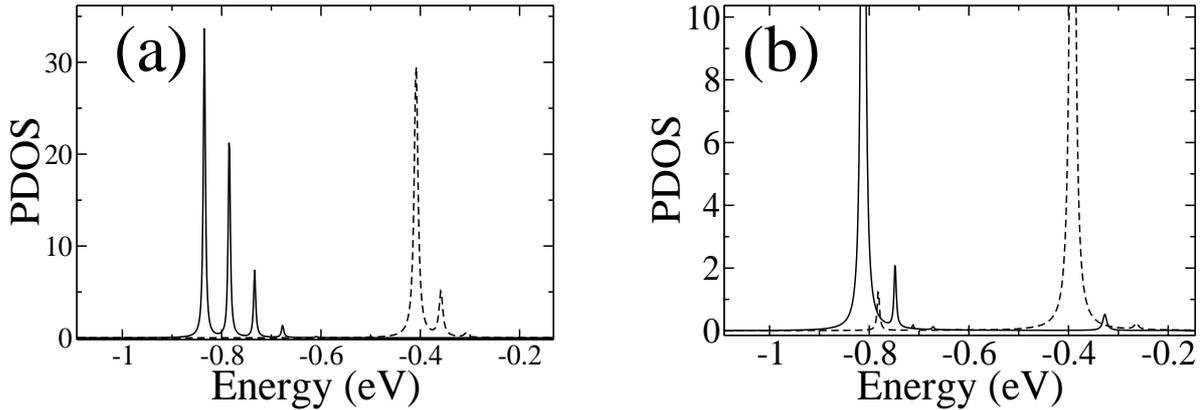}
\end{center}
\caption{
Projected density of states (PDOS) on molecular orbitals: bonding ($\sigma$)
in full line and antibonding ($\sigma^*$) in dashed lines.
The modes are considered separately in order to analyze the transmission
functions of the ABL mode (full line with dots in Fig.~\ref{ABL-CM})
and of the CM mode (dashed line in Fig.~\ref{ABL-CM}).
In (a) the  vibronic density of states
for the ABL mode is projected onto the two molecular orbitals and
in (b) the vibronic density of states of the CM mode. In (a) we
see two distinct series of vibronic peaks corresponding to the bonding
and the antibonding molecular orbitals since the ABL mode does
not couple these molecular orbitals. However (b) shows mixed structure
due to the coupling of molecular orbitals by the CM mode.}
\label{PDOSABL}
\end{figure}

In order to analyze the electron transmission, Fig.~\ref{ABL-CM},
it is more convenient to use projected density of states (PDOS)
of the full vibronic structure on molecular orbitals,
Fig.~\ref{PDOSABL}(a) for the ABL mode and (b) CM mode. 
These orbitals are linear 
combinations  of the two sites: 
\[
| \sigma \rangle  = \sqrt{\frac{1}{2}} ( | L \rangle + | R \rangle )
\]
and
\[
| \sigma^\star \rangle  = \sqrt{\frac{1}{2}} ( | L \rangle - | R \rangle )
\]
that diagonalize the uncoupled two-site Hamiltonian with eigenvalues
$\varepsilon_0-t_{\mbox{mol}}$ and $\varepsilon_0+t_{\mbox{mol}}$, 
respectively, where $\varepsilon_0$
 is the level energy for each site and $t_{\mbox{mol}}$
is the hopping matrix element between sites.

In this new basis set, the above coupling matrices, Eq.~(\ref{CM}) and (\ref{ABL}),
become:
\begin{equation}
{\bf M}_{CM}=\left[ \begin{array}{cc}
0 & m_3 \\
m_3 & 0 \end{array} \right],
\label{CMmo}
\end{equation}
and 
\begin{equation}
{\bf M}_{ABL}=\left[ \begin{array}{cc}
m_1+m_2 & 0 \\
0 & m_1- m_2 \end{array} \right].
\label{ABLmo}
\end{equation}
that hint at the interpretation of the above vibronic peaks. In the case of the CM mode,
the $\sigma^\star$ and  $\sigma$ orbitals are coupled via the vibration, while in the ABL case, the molecular
orbitals are not mixed by the electron-vibration interaction. This case can then
be interpreted as two single-sites connected to a vibration, and hence
two vibronic sequences are associated with the spectral feature of
$\sigma^\star$ at           and the spectral feature of $\sigma$ at   . Since, the effective
electron vibration coupling is $m_1+ m_2$ for $\sigma$, the number of peaks and the general
vibronic sequence corresponds to stronger coupling than the vibronic sequence
of the $\sigma^\star$ peak. This is clearly seen in the PDOS on $\sigma^\star$ and $\sigma$, Fig.~\ref{PDOSABL}, where
we see that the vibronic sequences with $\sigma^\star$ and $\sigma$ character 
are energetically localized near the original molecular-orbital peaks.

The CM mode mixes $\sigma^\star$ and $\sigma$, hence we obtain vibronic peaks 
that are shared in the PDOS
over both molecular orbitals. 

%Sergio - Semana del 25/02/08

\begin{figure}[tb]
\begin{center}
\includegraphics[width=0.95\columnwidth]{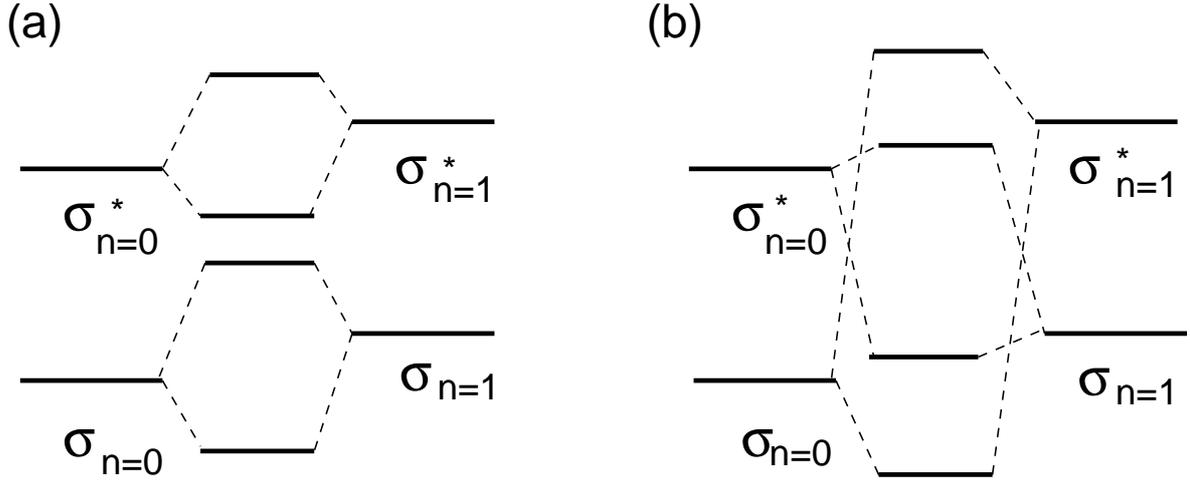}
\end{center}
\caption{Hybridization scheme for ABL mode (a) and for CM mode (b) considering only the two first 
vibrational levels $n=0$ and $n=1$. The levels
on the left and right are without electron-vibration coupling, but the right
levels are shifted by one quantum of vibration with respect to the
left levels. The coupled orbitals have a different vibrational state $n$ 
because they are mixed by the electron-phonon interaction.
It is clearly seen that in the ABL case, the coupling does not mix
the molecular orbitals between them ($\sigma_{n=0}$-$\sigma_{n=1}$,
$\sigma^{\star}_{n=0}$-$\sigma^{\star}_{n=1}$),
while in the CM case, the orbitals are mixed ($\sigma_{n=0}$-$\sigma^{\star}_{n=1}$,
$\sigma^{\star}_{n=0}$-$\sigma_{n=1}$). 
}\label{scheme}
\end{figure}

%%%%% Fin

For small enough couplings, we can estimate the vibronic structure keeping just $n=1$ in the
vibrational part. Hence we can diagonalize Hamiltonian~(\ref{Hamiltonian})
for the sites connected to the vibrations. 
This can also be expressed by a hybridization scheme, Fig.~(\ref{scheme}). We realize
that the new peaks correspond one-to-one to the peaks found in the PDOS, and that
they have contributions in different ratio from the $\sigma^\star$ and $\sigma$ orbitals according to
the matrix elements of {\bf M}. Depending on the magnitude of 
$g_M=(\frac{m_3}{\Omega_{CM}})^2$
more or less phonons, $n$, will be needed to converge the vibronic 
sequence~\cite{Mahan}, and more or less
elements will be included in scheme~(\ref{scheme}). In order to obtain the type of
diagram (a) or (b), only knowledge of the symmetry of the system is required.

In the present case, while the CM mode mixes $\sigma^\star$ and $\sigma$ orbitals efficiently,
the ABL mode does not mix them. Hence, only this last case can be understood by a 
single-level model.

%Time-dependence

It is interesting to study the time dependence of the electron occupation
of both sites depending on the considered mode.  Given the ``shuttle-like'' motion
of the CM mode, one could expect that both sites populate at the
same time when they approach together the left electrode
and then depopulate when they approach the right electrode.
On the other hand,
the ABL motion is a stretch mode and hence, one site would
populate while the other one depopulates.

\begin{figure}[tb]
\begin{center}
\includegraphics[width=1\columnwidth]{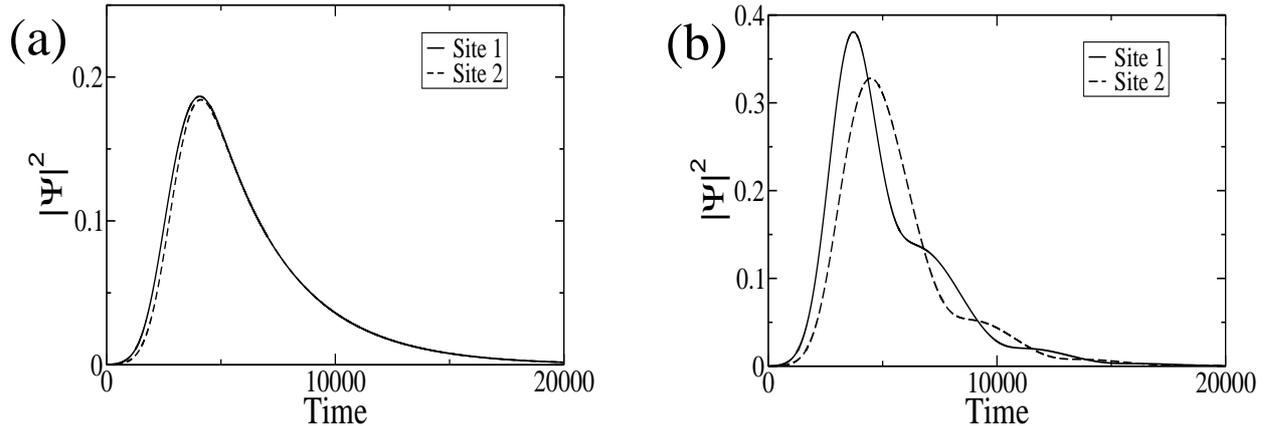}
\end{center}
\caption{
Population of the ABL mode (a) and of the CM mode (b) on the two
different sites that are connected to the vibration as a function
of time (atomic units). The ABL mode shows the same population for
the two sites as a function of time, while the CM mode presents
a time dependent sequential population.
}\label{Pop2}
\end{figure}

However, in Fig.~\ref{Pop2} (a) we see that the ABL mode leads to population
of both sites at the same time, while in presence of the CM mode, Fig.~\ref{Pop2} (b), the
population sequence is shifted and dependent on time.
In this case, the wave packet has been centered about the energy
$\varepsilon_{\sigma}$ and its energy span is smaller than the difference
between $\varepsilon_{\sigma^\star}$ and $\varepsilon_{\sigma}$. The reason for
this behavior is that one should think in terms of molecular orbitals
rather than site orbitals. In this case, the ABL motion does not couple
the $\sigma^\star$ and $\sigma$ orbitals, hence, the electron populates one of the molecular
orbitals, $\sigma^\star$,
at one time and interacts with the vibration in the same way as
in the case of the single site. Instead of a single site, 
we have a single level,
otherwise there is no physical difference.
Let us approximate the wave packet by the evolution on the two sites:
\[ |\Psi(t)_{\mbox{ABL}}|^2 \approx |\psi_{\sigma}|^2\]
without any time dependence (except the one due to the coupling to the
electrodes that we have neglected in this simplified discussion) because
$\psi_{\sigma}$ is basically an eigenstate of the Hamiltonian.
This explains Fig.~\ref{Pop2} (a).

In the case of the CM mode, we saw above that a single-site analogy cannot be applied
and we have the two molecular orbitals involved in the wave packet propagation.
Let us again approximate the wave packet by the evolution on the two sites:
\[ |\Psi(t)_{\mbox{CM}}|^2 \approx |\psi_{\sigma^\star}|^2 + |\psi_\sigma|^2 + 2 Re (\psi^\star_{\sigma^\star} \psi_\sigma
 e^{i (\varepsilon_{\sigma^\star} - \varepsilon_\sigma) t} )\]
where a clear time dependence subsists.
In good agreement with Fig.~\ref{Pop2} (b).

\subsection{Two-sites and two vibrations}

When more than one phonon per mode is needed to be converged with 
respect to the electron-vibration coupling, the inelastic transmission
cannot be separated in additive contributions from different modes and
all of them have to be considered at the same time.

In the above case, the truncated Hamiltonian reads now:
\[
\tiny{\left(
\begin{array}{cccccccccccccccccccc}
 \mathcal{H}^{0} & \hat{M}_a & 0 &  \hat{M}_c & 0  \\
 \hat{M}_a &  \mathcal{H}^{0}  + \hat{\Omega}_a &\sqrt{2} \hat{M}_a & 0 &\hat{M}_c & 0  \\
 0 &\sqrt{2}  \hat{M}_a &  \mathcal{H}^{0}  +2\hat{\Omega}_a & 0 & 0 &  \hat{M}_c & 0  \\
 \hat{M}_c & 0 & 0 &  \mathcal{H}^{0}  +\hat{\Omega}_c & \hat{M}_a & 0 &  \sqrt{2} \hat{M}_c & 0 \\
 0 & \hat{M}_c & 0 & \hat{M}_a &  \mathcal{H}^{0}  +\hat{\Omega}_c+\hat{\Omega}_a  &\sqrt{2}  \hat{M}_a & 0 & \sqrt{2} \hat{M}_c & 0   \\
 & 0 & \hat{M}_c & 0 &\sqrt{2}  \hat{M}_a &   \mathcal{H}^{0}  +\hat{\Omega}_c+ 2\hat{\Omega}_a &0 & 0 & \sqrt{2} \hat{M}_c \\
 & & 0 & \sqrt{2} \hat{M}_c  & 0 & 0 &  \mathcal{H}^{0}  +2\hat{\Omega}_c & \hat{M}_a & 0  \\
 & & & 0 & \sqrt{2} \hat{M}_c & 0 & \hat{M}_a &  \mathcal{H}^{0}  +2\hat{\Omega}_c+\hat{\Omega}_a & \sqrt{2} \hat{M}_a    \\
 & & & & 0 & \sqrt{2} \hat{M}_c & 0 & \sqrt{2} \hat{M}_a &  \mathcal{H}^{0}  +2\hat{\Omega}_c+2 \hat{\Omega}_a   . 
 \end{array}
 \right)}
\]

Here the subindex $a$ refers to the ABL mode and $c$ to the CM one.
At intermediate coupling, we need some 3 phonons to converge the transmission. Hence,
the Hamiltonian matrix in this basis set is composed of $9 \times 9$ blocks, each
block contains the infinite number of electron sites, that we deal with
as above.

The inclusion of several modes changes the transmission from the superposition of
peaks coming from each mode because the spectral weights change and the peaks shift.
This is seen in Fig.~\ref{ABL-CM}. There the peaks of the full transmission are shifted with
respect to the sum of peaks from the CM and ABL modes. This can be understood
in terms of the different Hilbert space that is considered as new modes are included.
Indeed, since the convergency in phonons, $n$, means the shift of peaks,
the transmission also needs to be converged with respect to modes.
In terms of Green's function
this is easily understandable because there are terms in the self-energy
stemming from all modes.

%Describir un caso con detalle

We can improve our previous hybridization diagram to describe the peak structure, by including
all modes.
This type of correlation diagram has been
termed
``progression of progressions'' and
 used in the literature to explain the vibronic sequence of C$_{60}$ and naphtalocyanine
molecules~\cite{Ho2005,Ho2007}. One word of caution is important: 
the knowledge of the
actual coupling matrices is needed in order to perform the correct hybridization of orbitals.
The hybridization scheme can be obtained by symmetry arguments alone, however the
separation and 
strength of the vibronic peaks need a quantitative evaluation of the matrix 
elements, which is increasingly difficult with the number of involved modes.

\section{The meaning of dips and peaks in the \d2}

The increase or decrease of conductance over the vibrational threshold has been used
to determine the occurrence of vibrational excitation during electron flow through 
molecular electronic states~\cite{Stipe,Agrait}. 

A simple explanation for the increase
of conductance in the tunneling regime was already advanced
 in IETS of metal-insulator-metal
junctions~\cite{Hansma}. The conductance increases because new conduction channels
become available above a vibrational threshold. This interpretation is correct
when the vibrational side bands of the electron transmission lie beyond $\hbar \Omega$
of the electrode's Fermi energy. As we have already seen, opening a new channel means
that we have to add the $n=1$ contribution to the $n=0$ contribution of the transmission.
Hence, the \d2 will present a positive peak at positive voltage, and negative at
negative voltage~\cite{Hansma}.

However, we can also have decreases of conductance. This is experimentally found in 
tunneling~\cite{Ho_O2} and in contact regimes~\cite{Agrait}. 
In the present theory, one always adds a positive contribution to the transmission
above the vibrational threshold, but the change in conductance is given by
the slope of the transmission, and this can have a rapid variation in the presence
of a sideband. In the case where the sideband is in the tail of the main peak,
and the width of the main peak is in the order of the vibrational frequency, then
the slope will change from a smoothly varying one to the faster varying sideband slope,
giving rise to a negative \d2 at positive voltage. 
Let us give two examples of the above cases.

\subsection{Tunneling regime}
 % casos de CO y de O_2

\begin{figure}[tb]
\begin{center}
\includegraphics[width=0.65\columnwidth]{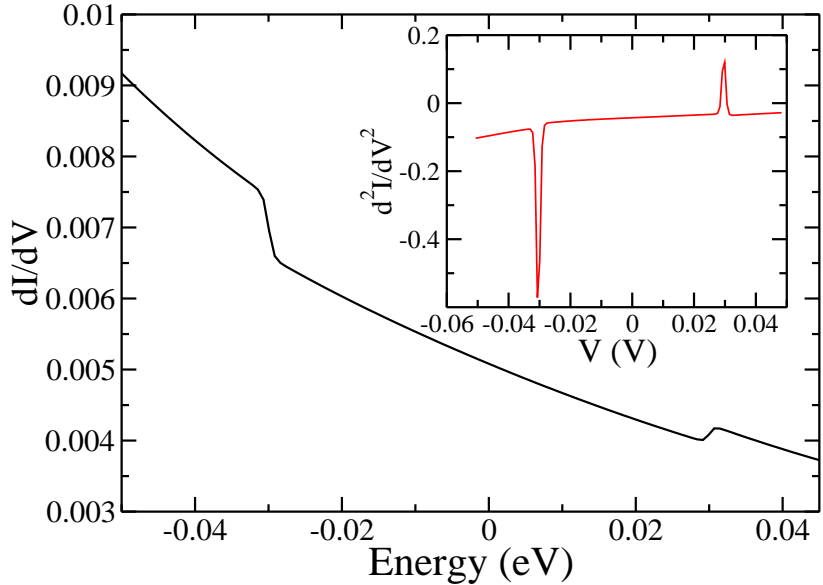}
\caption{
Conductance for a model system of two levels of CO on Cu(100). In the inset
the \d2 is depicted. The asymmetry in the size of the 
change in conductance is a consequence of the two-level model.
The variation in the conductance away from
 the vibrational threshold is due 
to the energy dependence of the two-level density of states. As can be seen
in the \d2, the variation is small with respect to the inelastic
change in conductance.
}\label{CO}
\end{center}
\end{figure}

In the case of the excitation of the CO frustrated rotation mode~\cite{Lauhon}, the
$2 \pi^*$ and the $ 5 \sigma$ molecular orbitals are coupled via the electron-vibration
coupling. This is easily seen by symmetry arguments~\cite{Lorente2005}, since the mode is antisymmetric
with respect to the planes containing the molecular axis and the non-zero matrix elements
will couple a symmetric orbital ($5 \sigma$) with an antisymmetric one (the $2 \pi^*$)~\cite{Lorente2000b}.
Assuming weak coupling, we can neglect the effect of all other modes and estimate the change
in conductance by using the two-site model coupled to a vibration that we presented above. Figure~\ref{CO}
shows the result of the conductance, Eq.~(\ref{conductance}), and in the corresponding \d2. We see that the Fermi
energy plus $\Omega$ for the frustrated rotation, gives the threshold where the transmission for
$n=1$ enters, giving a clear discontinuity in the conductance (positive contribution) and hence
a positive peak in the \d2. This is the case in metal-insulator-metal
junctions~\cite{Hansma}. The two-level model
leads to different inelastic efficiencies at positive and negative voltage bias.

\begin{figure}[tb]
\begin{center}
\includegraphics[width=0.75\columnwidth]{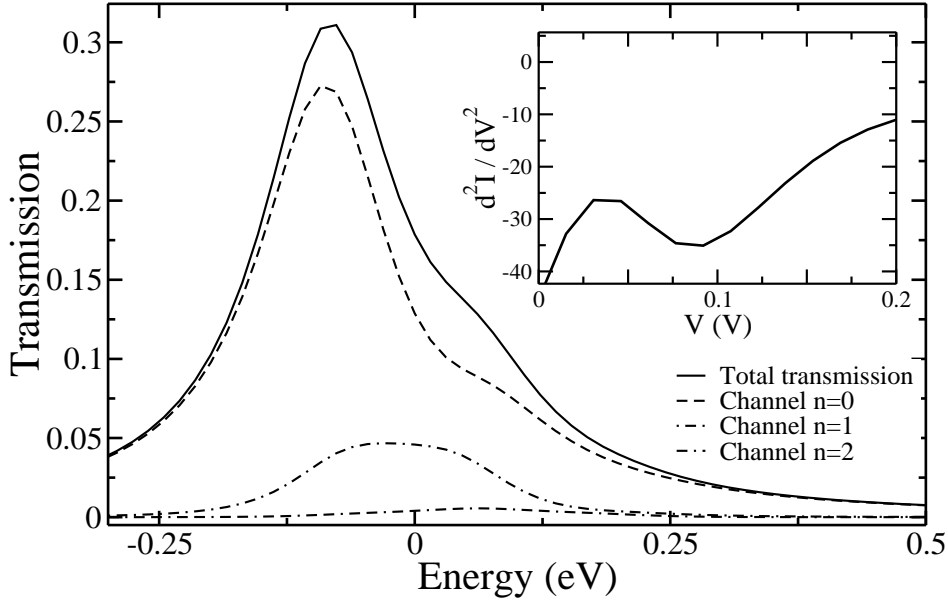}
\caption{
Tranmission for a model system of two levels of O$_2$ on Ag (110). In the inset
the \d2 is depicted. 
The decrease in the \d2 originates in the change of slope of the transmission
due to the vibrational side bands. These side bands belong to all vibrational channels,
in particular to the elastic one. Hence, the contribution of the elastic channel alone gives
a decrease in conductance because the density of states is rapidly changing due to the
electron-vibration coupling.
}\label{O2}
\end{center}
\end{figure}

However, the case of IETS of O$_2$ on Ag (110)~\cite{Ho_O2} is different. Experimentally, dips are
found instead of positive peaks in the \d2. Let us consider the excitation of the internal
stretch mode. By computing the PDOS on molecular orbitals~\cite{Olsson}, we know that the
$\pi_g$ perpendicular to the surface is close to the Fermi energy, and that tunneling
takes place through it~\cite{Olsson}. The mode has the same symmetry as the molecular orbital
and the diagonal matrix element will be different from zero. We are hence in a 
case that can be approximated by the single-site model. Figure~\ref{O2} shows the result. We have
located the $\pi_g$ at the Fermi energy as the PDOS on molecular orbitals~\cite{Olsson} seems to
indicate. The orbital width is taken as  0.1 eV  
  and the frequency energy is 0.08 eV.
We see that the transmission function is an asymmetric peak because the sideband is inserted
in the tail of the main peak. We also include the contribution of $n=1$ above the Fermi energy plus
the frequency in order to calculate the \d2 , as for the CO case
while we prefer to plot the complete one-electron transmission
because in the present case the initial level occupation matters and
we cannot treat it in this one-electron approach. 
Instead, we realize that the see the change of slope due
to the side bands in the $n=0$ and $n=1$ contributions. The \d2 gives a dip corresponding
to the change of slope of the $n=0$ contribution. Hence, the elastic component of the transmission
contains the information of the IETS. The decrease of conductance in this case is
due to the rapid change of transmission already in the elastic channel. It is then
a decrease due to the variation of the vibronic density of states.

From these models we conclude that negative peaks in the \d2 are due to sidebands in the elastic
transmission. This statement is equivalent to the one found in the pioneering work by Davis~\cite{Davis}.
Davis realized that a decrease of conductance could be found in some especial circumstances. Namely,
that the single-site resonance was at the Fermi energy and that the resonance width was of the order
of the vibrational frequency. These are the same conditions as we find. The interpretation by Davis
is that virtual phonons were emitted and absorbed giving rise to an interference pattern. 
This interpretation is based on perturbation theory, and it just accounts for the vibration's effect
on the elastic channel. In our terms, it is just the vibrational sideband of the elastic channel. 
The only cases in which
a vibrational sideband can yield a decrease in conductance is when the main peak of the
transmission is at the Fermi energy~\cite{Persson}
 (this is half-filling, we discuss below that we cannot
have half-filling in our models), and the electronic widths are of the order than the vibrational
frequency, so as to enlargened and distort the main peak by the sideband. In the case of 
half-filling, the sidebands are symmetric with respect to the Fermi energy~\cite{Hyldgaard}
giving rise to a peak at negative voltage and to a dip at positive one in the \d2. 
The voltage of the dip does not exactly correspond to the 
vibrational frequency, since it correspond to the largest slope
of the vibrational side band, not its maximum. Hence, these results suggest
that the voltages at the dips are not direct measurements of the vibrational
frequencies, while in the case of the voltages corresponding to
 peaks in the \d2, are direct readings of the vibrational frequencies.

There is certain confusion in the literature because perturbation theory is currently used. In this case the current in the absence of electron-phonon coupling
is confused with the elastic current. 
As can be seen in Fig.~\ref{Transmission} (a) in the absence of electron-phonon
coupling one gets a single lorentzian peak (dashed lines), however
the elastic contribution is the $n=0$ curve of Fig.~\ref{Transmission} (b). Hence, it is wrong to identify the elastic current
with the one without electron-vibration coupling.

\subsection{Contact regime}
%comparison with Magnus

Paulsson and co-workers~\cite{Magnus} have shown that in the case of symmetric
contact to the electrodes one can continuously pass from a peak to a dip in the \d2 by
increasing the transmission probability. At transmission 1/2 the threshold between both
behaviors is found. In the case of contact, a large density of states is pinned
at the Fermi energy. Hence, it is $\Gamma$  the parameter that controls the
coupling to the leads and, therefore, the passage from peaks to dips in the
change of conductance at the vibration threshold.
Recent experimental evidence has been reported in \onlinecite{Tal}.

\begin{figure}[tb]
\begin{center}
\includegraphics[width=0.75\columnwidth]{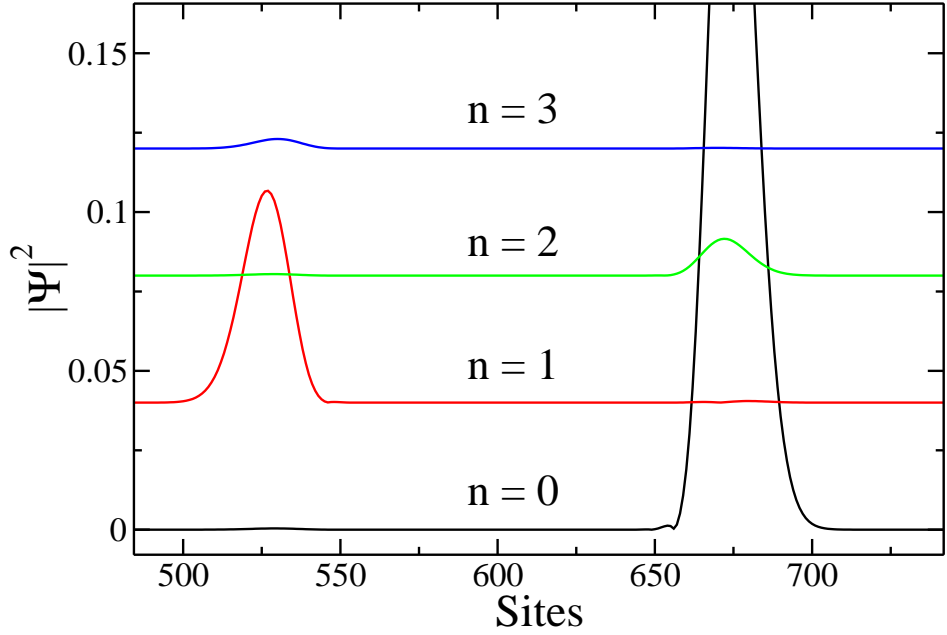}
\end{center}
\caption{ (Color online)
Electronic wave packets for different vibrational channels at a time instance
after the elastic wave packet has passed the vibrating chain centered
at site 600. Four electronic
sites are connected to an ABL vibration. This 4-site chain is strongly
coupled to the leads, representing the contact regime. Hence, the
elastic wave packet is basically unperturbed by the chain, but the vibrationally
excited channels have wave packets that are reflected for odd $n$, and
transmitted for even $n$.
}\label{Cuatro_n}
\end{figure}

Figure~\ref{Cuatro_n} shows the behavior of the electronic wave packet after
collision with a 4-site chain in the contact regime. The parameters are
those from Ref.~\onlinecite{Thomas2004}. The chosen mode is the ABL, that
corresponds to the one giving the largest change in conductance in
gold monoatomic chains~\cite{Agrait}. The wave packet is mainly
transmitted in the elastic channel as is to be expected from the contact
regime, however the vibrationally excited wave packets are mainly
reflected in agreement with the interpretation of Ref.~\onlinecite{Agrait}.
However, the behavior is more complex: electronic wave packets
in channels with odd $n$ are reflected, even $n$ are
transmitted. In this particular calculation, the reflection of the $n=1$ component
is due to finite-band effects. As we discussed above, a finite band leads to the
closing of transmission channels
when inelastic effets are operative,
 increasing the electronic wave packet reflection. As the number of
vibrational excitations increases, the closing of channels corresponds to
each excitation process, implying an increase of the reflection in each case. As
a consequence, even excitations mean even number of reflections, leading to
an increase in transmission. Indeed, we have found this behavior for large transmission
 even in the single-site case. 

Some of the calculations of Ref.~\onlinecite{Magnus} have been 
performed in the wide-band
approximation, and hence the closing of channels
due to finite-band effects is not present. Hence,
 we cannot conclude that the decrease in conductance leads
to an increase in electron reflection such as the
one found in this section. Instead, we think their calculations correspond
to the regime where rapid changes in the vibronic density of states are found
which correspond to the precedent section.

\section{Many-electron and electron-electron efects in the presence of vibrations}

Our calculations are exact regarding the vibrational dynamics during 
electron transport. The price to pay is the neglect of multielectron
effects. The description of multielectron effects together with vibrational
excitation necessarily implies approximations. A recent treatment of both
effects is the 
one by Galperin et al.~\cite{Galperin2006}. Despite approximations, their treatment is
correctly evaluating the vibrational dynamics. Indeed, we can reproduce their
results,  and the
main difference  between both calculations
 is an absorption shoulder appearing below the first
peak in the multielectron case.  According to Ref.~\onlinecite{Galperin2006} 
this is due to the propagation of holes coexisting with electron transport.
 This is absent in our one-electron treatment.

Even in the absence of a direct electron-electron interaction in the
Fr\"ohlich-Holstein model used here, Eq.~(\ref{Hamiltonian}), the electron-vibration coupling
can lead to effective electron-electron interactions, see
for example Ref.~\onlinecite{Hyldgaard}. 
This leads to electron-electron correlations and has extensively been studied
in the literature in the so-called bi-polaron problem~\cite{Zhang}. 
This is absent in our
one-electron approach, but could in principle be treated by using two-electron propagations.

Perhaps, more important for the description of 
transport in metallic systems is the
absence of the initial electron occupation in the present approach. 
This is easily
solved by usual non-equilibrium Green's functions approaches~\cite{Hyldgaard} 
at the price
of simplifying the treatment on the phononic description. 
Indeed, while the present
approach treats the nuclear-coordinate description exactly within 
the one-electron
problem and the harmonic approximation, 
NEGF approaches need to rely on approximate
treatments~\cite{Galperin2006,Hyldgaard,Mitra}.
This situation can be solved by using a time-dependent Hartree-Fock approach
to generate a many-body wave packet, extending the present treatment
to a multielectron case. Some encouraging results in time-dependent
Hartree-Fock and beyond techniques have been presented in
 Refs.~\onlinecite{Nest,Krause}.

\section{Summary and conclusions}

Time-dependent wave packet propagations for the study of 
inelastic effects in electron transport can be very interesting
because of the physical picture they permit us to develop
as well as the good size-scaling properties that they
can have. Indeed, in the case of sparse Hamiltonians, the
scaling is basically linear with the system size, when
Lanczos-like propagation methods are used.

In the present work we have analyzed the electron-vibration
problem when a single electronic site is connected to
two electrodes. We have used a time-dependent description
to understand the vibrational excitation sequence, the
electronic phase shift and the interference patterns.
The time-dependent calculation has given us access to
the same quantities we can calculate using an energy-resolved
theory, but with the time-dependent insight.

We have also studied the case of two sites and two vibrational
modes, and we have realized that the two-site problem can be simplified to 
the single-site one when molecular orbitals can be used. The electron-vibration
coupling will determine when molecular orbitals are meaningful. This is
seen by changing the electron-vibration coupling matrix to the
molecular-orbital basis set. Pursuing this idea, we have developed
a simple hybridisation scheme that permits us to understand the 
vibronic structure of an electron transmission function in terms
of the symmetry of the electron-vibration couplings. 
We have shown, that for
medium and strong electron-vibration coupling all modes need to be considered.

We have applied our simple model to get some insight in the case of the
vibrational excitation of CO on Cu(100)~\cite{Lauhon} and 
O$_2$ on Ag(110)~\cite{Ho_O2}. Due to our one-electron treatment, the
calculations are missing fundamental ingredients, but they
permit us to explain the IETS signals in terms of vibrational channels.
In the case of O$_2$, the calculation show that the elastic channel
leads to a decrease in the \d2. Contrary
to perturbation theory, the elastic channel is not the electronic
structure in the absence of electron-vibration coupling, but the $n=0$
transmission function. Hence the elastic channel contains all of
the information about the electron-vibration coupling. In particular,
the elastic contribution will contain vibrational satellites. These
satellites or side bands have a strong energy dependence that naturally
leads to dips in the \d2.
Indeed this means that the density of states that should be considered
is the vibronic one, this density of states contain rapid variations
with the energy that lead to decreases of \d2.
In perturbation theory, one needs to consider the effect of the vibration
in the elastic channel to retrieve the effect of the vibration in
the electronic structure and hence to take into account the vibronic
density of states in the conductance.

We have also explored the case of contact with the two electrodes. In
this case, it has been shown that the conductance should drop
at the vibrational thresholds because the vibrations backscatter
the impinging electrons. Our calculation show that for
the first excited (and generally odd vibration channel, $n$) 
the electronic wave packet is
reflected at the vibration. However for even $n$ the wave packet is transmitted.
This behavior is due to the finite electronic band of the present calculation,
that leads to the closing of inelastic channels increasing reflection.
An even number of 
consecutive reflections lead to increased transmission.

New developments in time-dependent Hartree-Fock lead us to think
that a full time-dependent treatment of the physics usually explored
with non-equilibrium Green's functions will be soon
available permitting us to develop a different point of view
on inelastic effects in electron transport on the atomic scale.

\acknowledgements
Interesting discussions with 
A. Arnau, 
A.-G. Borisov, 
T. Frederiksen,
J.-P. Gauyacq, 
E. Jeckelmann,
T. Klamroth,
M. Paulsson, 
P. Saalfrank, and
H. Ueba  are gratefully
acknowledged.
N.L. acknowledges support from the Spanish MEC (No. FIS2006-12117-C04-01).
Computer resources of the Centre de Calcul de Midi-Pyr\'en\'ee (CALMIP) are
gratefully acknowledged.


\begin{thebibliography}{99}
\bibitem{Negative} We assume that the main difference between the negative
PES and the neutral one is the change in geometrical conformation of the
negative ion with respect to the neutral molecule. However, frequencies
also change and as a matter of fact the electron-vibration coupling
should be computed for the negative PES.
\end{thebibliography}
\end{document}